\documentclass[epsf,twocolumn,showpacs,preprintnumbers]{revtex4}
\usepackage{graphics}
\usepackage{graphicx}
\usepackage{dcolumn} 
\usepackage{bm}
\usepackage{epsfig}
\begin{document}
\title{Chemical Differences between K and Na in Alkali Cobaltates}
\author{K.-W. Lee and W. E. Pickett}
\affiliation{Department of Physics, University of California, Davis, 
            California 95616}
\date{\today}
\begin{abstract}
K$_x$CoO$_2$ shares many similarities with Na$_x$CoO$_2$, as well as
some important differences (no hydration-induced superconductivity has
been reported).
At $T_{c2}=20$ K, K$_{0.5}$CoO$_2$ becomes an insulator 
with a tiny optical gap as happens in Na$_{0.5}$CoO$_2$ at 52 K. This 
similarity, with a known common structure, enables direct comparisons
to be made.
Using the K-zigzag structure recently reported and the local density 
approximation, we compare and contrast these cobaltates at $x$=0.5.
Although the electronic structures are quite similar as expected, 
substantial differences 
are observed near the Fermi level.
These differences are found to be attributable mostly to the chemical,
rather than structural difference: although Na is normally considered
to be fully ion, K has somewhat more highly ionic 
character than does Na in these cobaltates.
\end{abstract}
\maketitle

\section{Introduction}
Takada {\it et al.} found superconductivity in the layered quasi-two
dimensional Na$_x$CoO$_2$ when intercalating enough water ($\sim 1.3$H$_2$O) 
to form a separate water layer between CoO$_2$ and Na layers.\cite{takada}
The nonsuperconducting dehydrated Na$_x$CoO$_2$ system shows a
rich phase diagram, which significantly depends on $x$.\cite{foo}
For $x <0.5$, the system shows weakly correlated band-like behavior
including Pauli paramagnetism,
while the phase $x>0.5$ reveals correlated behavior such as 
large enhancement in linear specific coefficient, Curie-Weiss
susceptibility,\cite{prb04} and magnetic ordering for $x\geq 0.75$.

The most peculiar aspect of this system is an insulating phase 
at $x=0.5$\cite{foo} with a tiny gap $\sim 15$ meV.\cite{ncogap}
As the temperature is decreased, antiferromagnetic ordering of 
some Co spins appears at $T_{c1}=88$ K, and at $T_{c2}=52$ K 
there is a gap opening, which reflects the charge-ordering of
nonmagnetic $S=0$ Co1 ions 
and magnetic $S=\frac{1}{2}$ Co2 ions.\cite{mit,yokoi}
Using neutron diffraction studies, Williams {\it et al.} inferred 
the charge difference of 0.12$e$ between Co1 and Co2.\cite{cava2}
This value is much smaller than the 1$e$ value expected from 
a naive formal charge
concept, but is roughly consistent with the theoretically 
calculated value 0.2$e$ 
using a correlated band theory LDA+U method.\cite{prl05,prl06}
As a result, even though there is small charge difference between 
the Co ions,
the charge-disproportionation is accompanied by local moment formation
and the spins are consistent with the formal valences
Co$^{3+}$ and Co$^{4+}$.\cite{prl05,prl06}

The discovery of an unexpected insulating state
in Na$_{0.5}$CoO$_2$ (N$_{0.5}$CO) and hydration-induced
superconductivity has stimulated 
the study of isostructural and isovalent family A$_x$CoO$_2$ (A = K, Rb, Cs).
In spite of a few attempts to produce superconductivity in 
hydrated K$_x$CoO$_2$,
the amount of intercalated water is 0.8 or less,
forming only a monohydrate (K+H$_2$O) layer and no superconductivity has
been detected yet.\cite{fu,tang}

The K$_x$CoO$_2$ system has been known for three decades, since 
Hagenmuller and colleagues reported\cite{hagen1,hagen2} 
structure, transport, and magnetic
data on phases with $x$=1.0, 0.67, and 0.50. 
Recently, an insulating phase in K$_{0.5}$CoO$_2$ (K$_{0.5}$CO) 
has been studied in more detail by a few groups;\cite{watanabe,qian} 
Nakamura {\it et al.}\cite{nakamura} in the mid-1990s had reported
an almost temperature-independent resistivity well above a metallic value.
In  K$_{0.5}$CO, using NMR and neutron diffraction studies, 
Watanabe {\it et al.} observed similar temperature 
evolution as in N$_{0.5}$CO.\cite{watanabe}
At $T_{c1}=60$ K, a kink in the in-plane susceptibility $\chi_{ab}$
indicates onset of antiferromagnetic ordering.
The resistivity increases sharply at $T_{c2}=20$ K, 
signaling the charge-ordering. At this temperature, there is
an additional magnetic rearrangement, indicated by kinks in 
both $\chi_{ab}$ and $\chi_{c}$.
From $\mu^+$SR experiments Sugiyama {\it et al.} have obtained 
similar transition temperatures, 60 and 16 K, 
in metallic K$_{0.49}$CO.\cite{sugiyama}
The former is a magnetic ordering temperature 
from a paramagnetic state.
Based on a mean field treatment of a Hubbard model, 
they suggested there may be a linear spin density wave (SDW) state
between 16 and 60 K, while a commensurate helical SDW state exists 
below 16 K.
Additionally, K or Na ions order (structurally), resulting in formation of
a 2$\times$$\sqrt{3}$ supercell at T$_{c0}= 550$ and 470 K 
for K and Na ions, respectively.\cite{watanabe}
The tiny energy gap of similar magnitude with N$_{0.5}$CO has been
observed by Qian {\it et al.} with ARPES measurements.\cite{qian}

Several characteristics of N$_x$CO, in particular the superconductivity
upon hydration and effects of cation ordering, suggest that the behavior
in this system is sensitive to details of the electronic structure.
The fact that K$_{0.5}$CO is similar to N$_{0.5}$CO, yet shows clear
differences in behavior, indicates that a comparison of the electronic
structures of these systems
is warranted.
In this paper, we compare and contrast the two insulating systems
K$_{0.5}$CO and N$_{0.5}$CO.
Here correlation effects and detailed magnetic ordering are neglected,
but the observed $\sqrt{3}a_H\times 2a_H$ supercell including Na/K 
zigzag ordering is adopted. ($a_H$ is the hexagonal lattice constant.)

\begin{table}[bt]
\caption{Crystal structure comparison between K$_x$CoO$_2$ and Na$_x$CoO$_2$ 
  at $x$=0.5. The orthorhombic structures (space group: $Pmmn$, No. 59) 
  determined from Na- or K-zigzag ordering are a $\sqrt{3}a_H\times 2a_H$
  superstructure which is based on the hexagonal structure with a lattice 
  constant $a_H$.
  In this structure, the oxygens have three site symmetries, 
  two $4f$ and one $8g$.
  Here, $z_O$ is an oxygen height from the Co layers.
  A main difference in these structures is 
  that K$_{0.5}$CoO$_2$ has 12 \% larger $c$ lattice constant.
  The data are from Ref. \cite{cava2} for Na$_{0.5}$CoO$_2$
  and Ref. \cite{watanabe} for K$_{0.5}$CoO$_2$.
  }
\begin{center}
\begin{tabular}{cccc}\hline\hline
   parameters~~ &~~ $a_H$ (\AA)~~ &~~ $c$ (\AA)~~ & ~~~$z_O$ (\AA)~~~
   \\\hline
   Na$_{0.5}$CoO$_2$& 2.814 &11.06     & 0.971, 0.949, 0.983 \\
   K$_{0.5}$CoO$_2$ & 2.831 &12.50     & 0.965, 0.946, 0.981 \\\hline\hline
\end{tabular}
\end{center}
\label{table1}
\end{table}

\section{Crystal Structure and Calculation Method}
Although some aspects of the structure in the sodium cobaltates 
are still controversial (especially the alkali metal ordering),
all existing information for $x$=0.5 are based on the basic 
hexagonal structure. 
Recently, Watanabe {\it et al.} observed 
the orthorhombic $\sqrt{3}a_H\times 2a_H$ superstructure from a K-zigzag 
pattern for K$_{0.5}$CO.\cite{watanabe}
For comparison, we have used this orthorhombic structure for both 
cobaltates.\cite{cava1,cava2}
As shown in Table \ref{table1}, in this structure
the oxygens have three different site symmetries and slightly
different O heights (from the Co layers), leading to distorted
CoO$_6$ octahedra.
The averaged Co--O--Co bond angle is about 96.5$^\circ$ for K$_{0.5}$CO 
and 95.4$^\circ$ for N$_{0.5}$CO 
(this angle would be 90$^\circ$ for undistorted octahedra).
This distortion makes the three-fold $t_{2g}$ manifold split into
singlet $a_g$ and doublet $e_g^\prime$ bands.

The calculations reported here were carried out within the local density
approximation (LDA), using the full-potential 
local-orbital method (FPLO).\cite{fplo}
The basis sets were chosen as $(3s3p)4s4p3d$ for Co and K, $(2s2p)3s3p3d$ 
for Na, and $2s2p3d$ for O.
(The orbitals in parentheses denote semicore orbitals.)
The Brillouin zone was sampled with 98 irreducible $k$ points.

\begin{figure}[tbp]
\vskip 6mm
\resizebox{8cm}{6cm}{\includegraphics{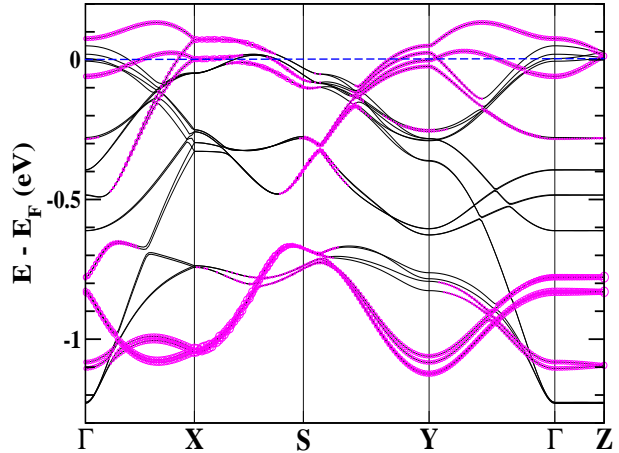}}
\caption{(Color online) Enlarged band structures of nonmagnetic
 K$_{0.5}$CoO$_2$ at the $t_{2g}$ manifold regime.
 The large $t_{2g}$-$e_g$ crystal field splitting of 2.5 eV makes 
 the $e_g$ manifold (not shown here) unimportant 
 for low energy excitations.
 The thickened (and colored) lines highlight bands having the strong
 Co $a_g$ character.
 The $S$ point is a zone boundary along $\langle110\rangle$
 direction. The horizontal dashed line indicates the Fermi energy $E_F$
 (set to zero).
 }
\label{band1}
\end{figure}

\begin{figure}[tbp]
\flushleft
\resizebox{3.8cm}{2.6cm}{\includegraphics{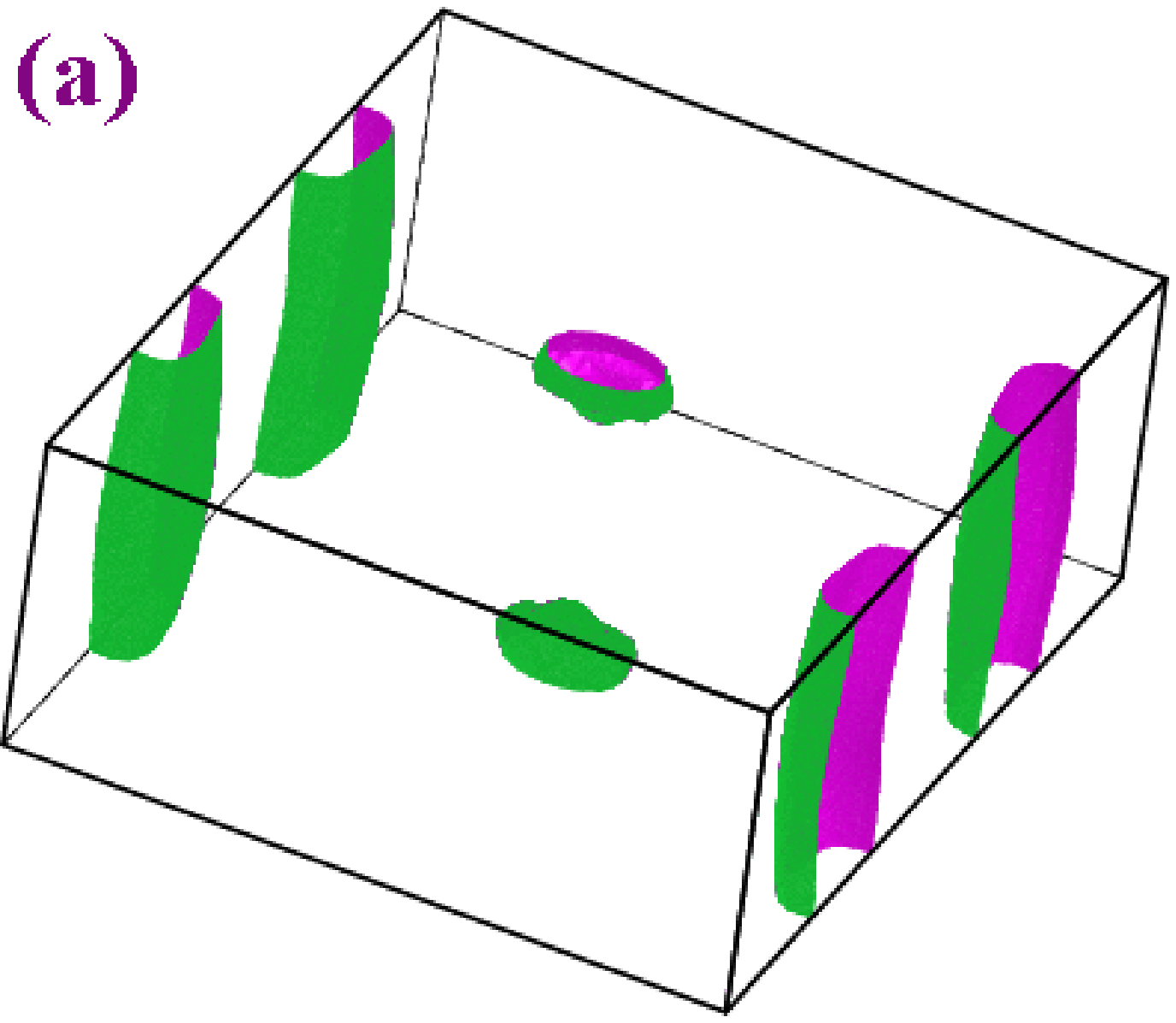}}
\flushright
\vskip -28mm
\resizebox{3.8cm}{2.6cm}{\includegraphics{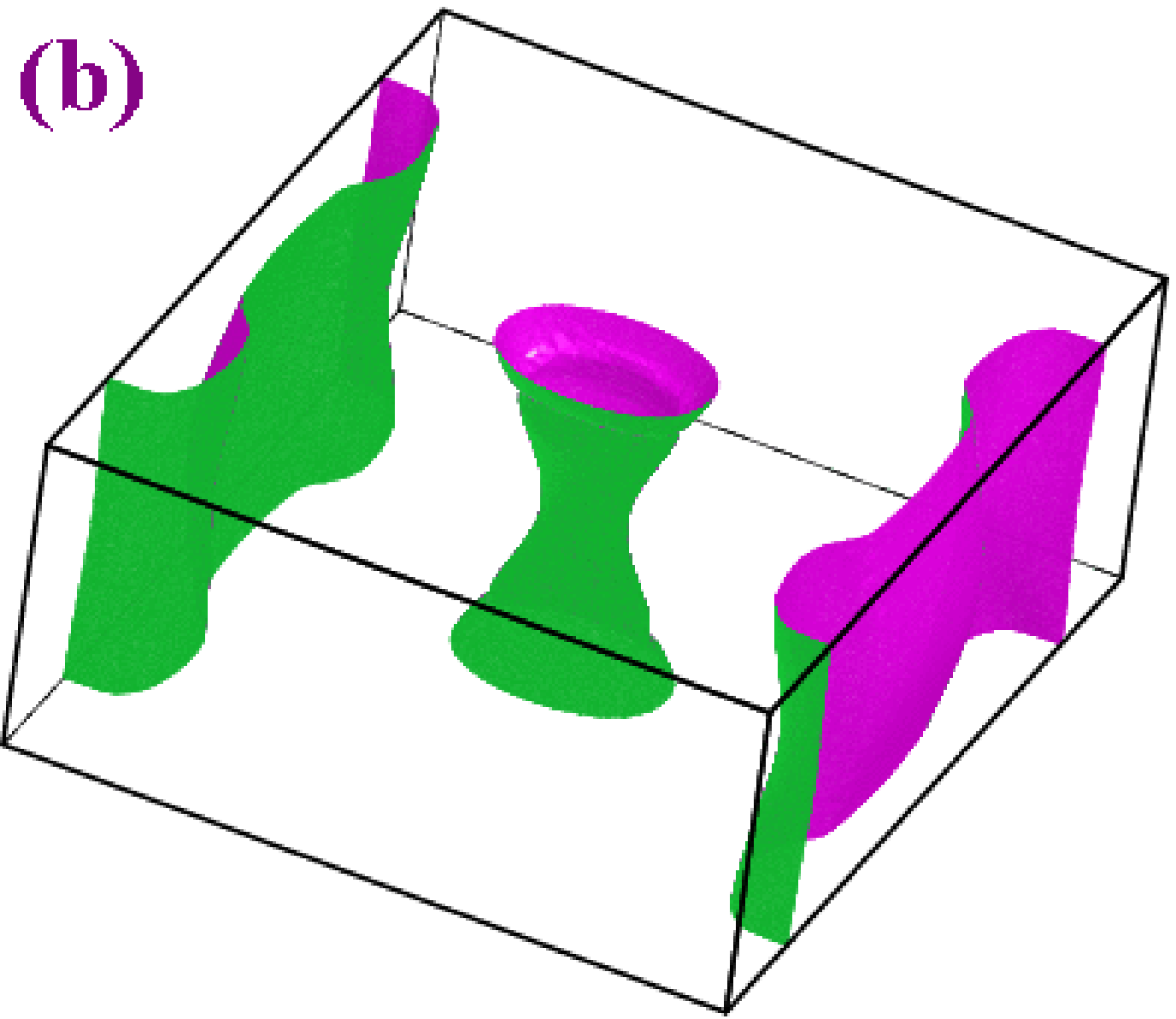}}
\vskip 1mm
\flushleft
\resizebox{3.8cm}{2.6cm}{\includegraphics{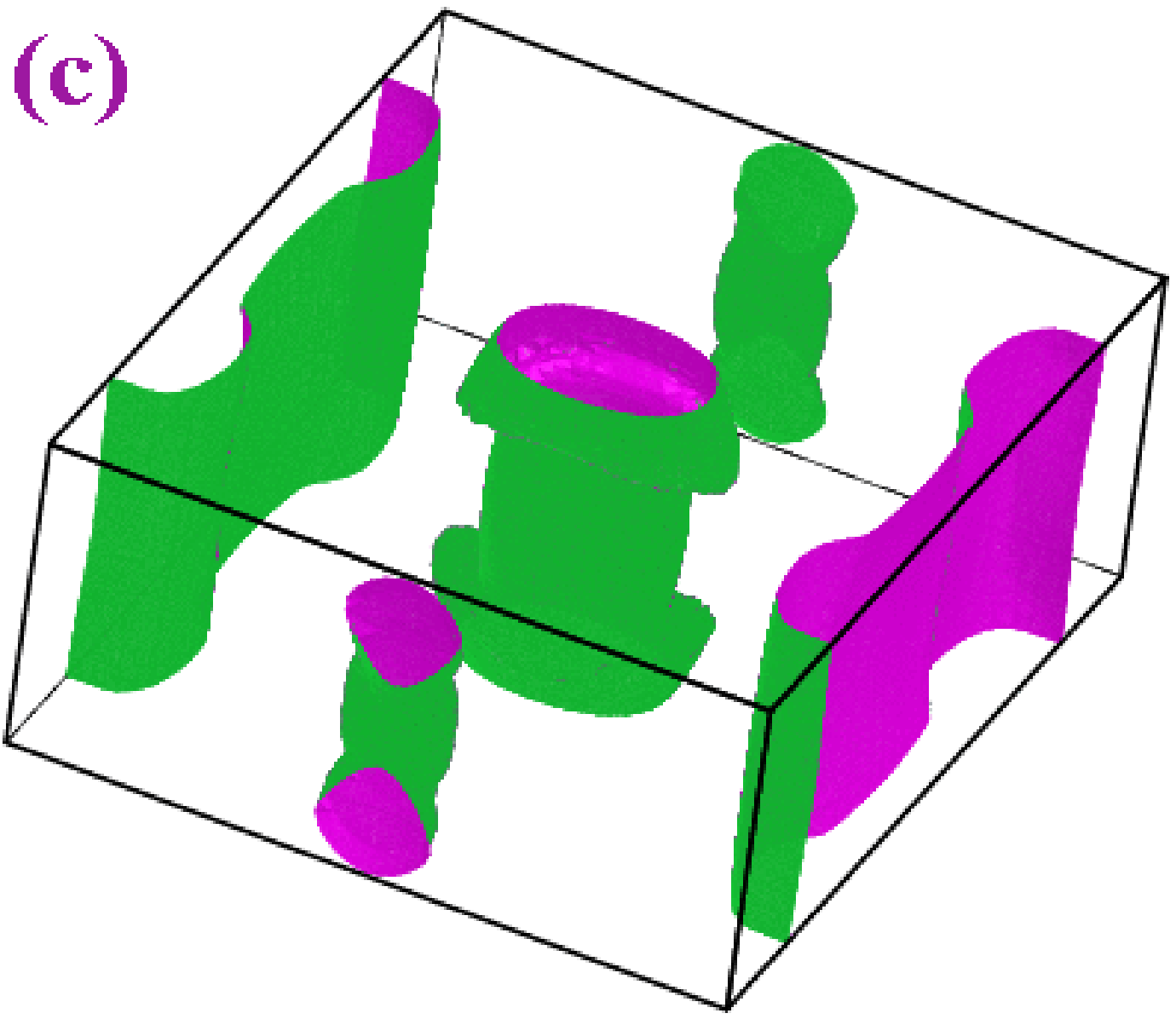}}
\flushright
\vskip -28mm
\resizebox{3.8cm}{2.6cm}{\includegraphics{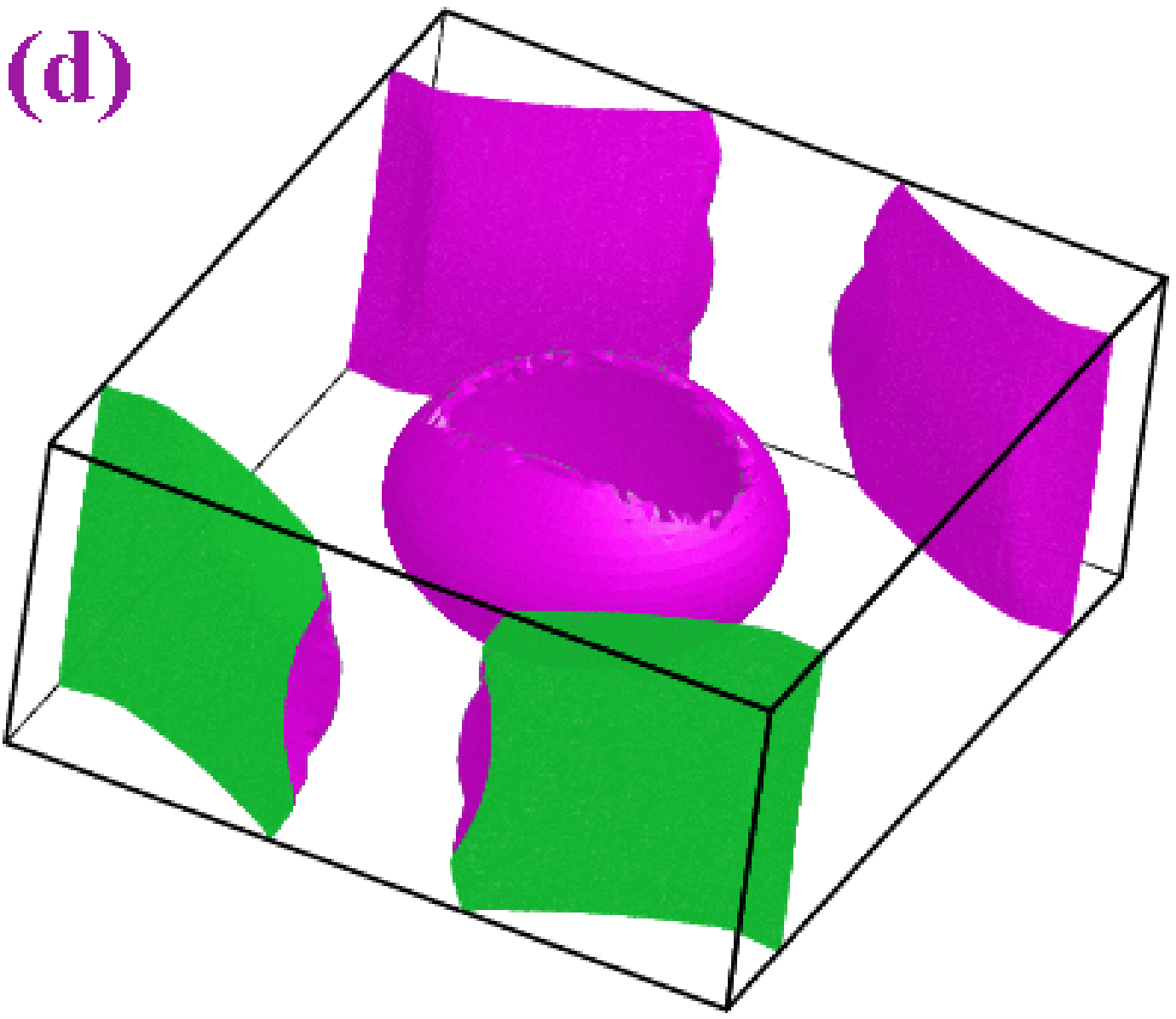}}
\caption{(Color online) Fermi surfaces of nonmagnetic K$_{0.5}$CoO$_2$,
 showing strong two-dimensionality.
 The band structure leads to six Fermi surfaces, but the first and sixth
 FSs are not shown here.
 The first FS is similar to (a), except a smaller cap
 at the Z points. The sixth FS has the same shape as (d), but it has no
 $\Gamma$ centered egg.
 The pink (darker) colored surfaces contain holes, whereas
 the green (lighter) colored surfaces hold electrons.
}
\label{FS}
\end{figure}

\section{Results}
\subsection{Magnetic energy}
In N$_x$CO, the FM state is generically favored energetically 
within LDA,\cite{singh00,prb04}
although this picture is physically correct only for 0.7 $< x <$ 0.9.
Our calculations show this tendency is also true for K$_x$CO.
The magnetization energy, defined by the energy difference
between nonmagnetic and ferromagnetic states, in N$_{0.5}$CO is 22 meV/Co, 
and the energy in K$_{0.5}$CO slightly increases to 26 meV/Co.
The small energy difference can be attributed to the higher magnetic moment
on Co in K$_{0.5}$CO, resulting from longer $c$ parameter in K$_{0.5}$CO. 
(This larger $c$ lattice constant results in increasing charge of 
each Co ion by 0.02$e$ in K$_{0.5}$CO, see below.)
From a simple Stoner picture, the small magnetization energy is consistent 
with small total magnetic moment of 0.5 $\mu_B$/Co.

\subsection{Electronic structure}
Now we will focus on the nonmagnetic state to understand the
microscopic chemical differences. 
As observed previously for all $x$ in N$_x$CO,\cite{singh00,prb04} 
the crystal field splitting between the partially occupied $t_{2g}$ 
manifold with 1.3 eV width and the unoccupied $e_g$ 
manifold with 1 eV width is 2.5 eV.
The large splitting makes the $e_g$ manifold irrelevant for low
energy considerations.

The band structure of the $t_{2g}$ manifold, showing 
strong two-dimensionality, is given in Fig. \ref{band1}. 
(This two-dimensionality is reflected in the Fermi surfaces displayed
in Fig. \ref{FS}.)
The $a_g$ character emphasized by the thickened (or colored)
lines is represented by the ``fatband" technique in Fig. \ref{band1}.
The $a_g$ character appears at both the bottom and top of
the $t_{2g}$ manifold, but the character is a little stronger in the
bottom.
This behavior is also observed in N$_{0.5}$CO.

\begin{figure}[tbp]
\vskip 6mm
\resizebox{8cm}{6cm}{\includegraphics{Fig3a.eps}}
\rotatebox{-90}{\resizebox{6.5cm}{7.5cm}{\includegraphics{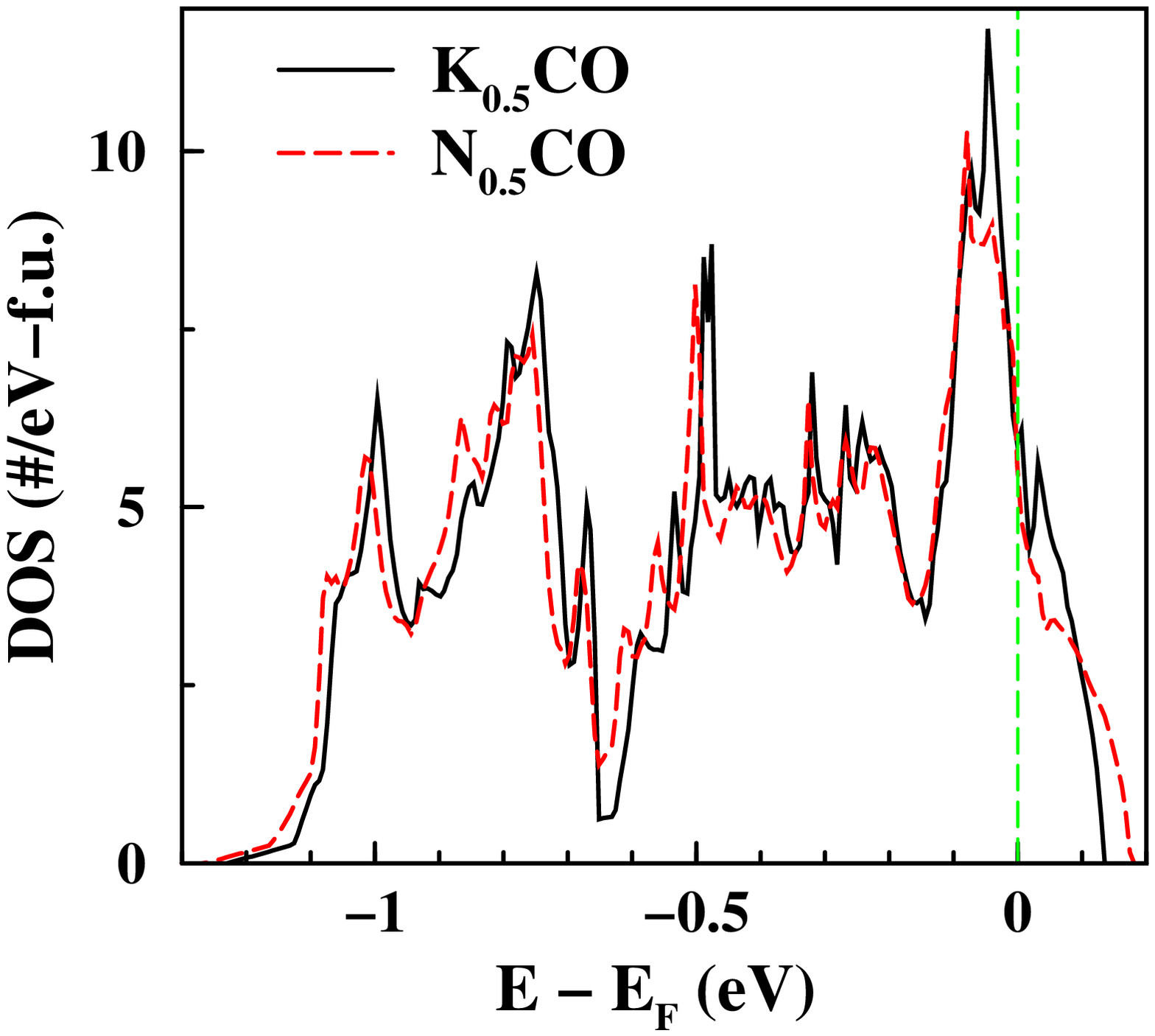}}}
\caption{(Color online) Comparison of electronic structure 
 between nonmagnetic K$_{0.5}$CoO$_2$ and Na$_{0.5}$CoO$_2$.
 Top: Enlarged band structures near $E_F$.
 Differences between the band structures are more noticeable at $E_F$,
 in particular at the $X$ and $Y$ points and along the
 $\Gamma$-$Z$ line.
 The band structure of K$_{0.5}$CO also shows much stronger 
 two-dimensionality.
 Bottom: Total densities of states per formula unit
 at the $t_{2g}$ manifold regime.
 K$_{0.5}$CoO$_2$ has about 10\% larger $N(0)$ than
 5.4 states/eV per a formula unit of Na$_{0.5}$CoO$_2$ (but invisible
 in this figure). Here, $N(0)$ is the density of states at $E_F$.
 The vertical dashed line denotes $E_F$.
 }
\label{band2}
\end{figure}

As expected from the larger $c$ lattice constant, 
K$_{0.5}$CO has a smaller bandwidth,
seen in both the O $p$ bands (not shown here) and Co $t_{2g}$ bands. 
The change in the bandwidth appears clearly at the top valence band 
in the enlarged band structures near $E_F$ depicted 
in the top panel of Fig. \ref{band2}.
The top valence band of  K$_{0.5}$CO has about 60 meV lower energy 
at the $\Gamma$ point and contains less holes,
leading to additional $E_F-$crossing valence band near the $X$ point
and along the $Y-\Gamma$ line.
This crossing produces additional Fermi surfaces 
of unfolded scroll-like shape
along the $X-S$ line, as displayed in (b) and (c) of Fig. \ref{FS}.
These Fermi surfaces are almost flat near the $X$ point, suggesting
enhancement of nesting effects.
These nesting effects would lead to SDW, suggested in K$_{0.49}$CO
by Sugiyama {\it et al}.
Absence of these Fermi surfaces in N$_{0.5}$CO 
may explain why SDW does not occur in the system.

An important distinction is the stronger two-dimensionality in K$_{0.5}$CO. 
At the $X$ and $Y$ points and along the $\Gamma$-$Z$ line, near $E_F$
there are nearly flat bands and saddle points in K$_{0.5}$CO. 
The bottom panel of Fig. \ref{band2} displays a comparison of the DOS  
of the two cobaltates in the $t_{2g}$ regime.
Strikingly, the Fermi energy (set to zero) of K$_{0.5}$CO lies 
midway between two sharp peaks at $-45$ and $35$ meV.
In addition, a van Hove singularity appears just above $E_F$ 
(at less than 10 meV).
These more complicated structures near $E_F$ lead to 
10\% higher DOS at $E_F$, suggesting an increased tendency toward
magnetic instability.

\begin{table*}[bt] 
\caption{Atom-decomposed charges, which are obtained from the Mullikan charge
 decomposition in the FPLO method, for each atom 
 in A$_{0.5}$CoO$_2$ (A=Na, K).
 The absolute numbers do not have a clear meaning, but differences
 reflect real distinctions in bonding.
 N$_{0.5}$CO$^*$ denotes Na$_{0.5}$CoO$_2$ with 
 the same crystal structure as K$_{0.5}$CoO$_2$.
  } 
\begin{center}
\begin{tabular}{ccccccccccccc}\hline\hline
 atom   &\multicolumn{3}{c}{A} & &\multicolumn{3}{c}{Co} & &\multicolumn{4}{c}{O}
    \\\cline{2-4}\cline{6-8}\cline{10-13}
site label  & $2a$ & $2b$ & $Ave.$ & & $4f$ & $4d$ & $Ave.$ & 
           & $4f$ & $4f$ & $8g$ & $Ave.$ \\\hline
 K$_{0.5}$CO & +0.72 & +0.68 & +0.70 & & +1.58 & +1.60 & +1.59 & 
             & $-0.97$ & $-0.97$ & $-0.97$ & $-0.97$ \\
 N$_{0.5}$CO & +0.64 & +0.63 & +0.63 & & +1.60 & +1.62 & +1.61 &
             & $-0.94$ & $-0.97$ & $-0.97$ & $-0.96$ \\
N$_{0.5}$CO$^*$ & +0.64 & +0.61 & +0.63 & & +1.59 & +1.60 & +1.59 &
             & $-0.94$ & $-0.95$ & $-0.96$ & $-0.95$ \\\hline\hline
\end{tabular}
\end{center}
\label{table2}
\end{table*}

\begin{figure}[tbp]
\vskip 6mm
\resizebox{8cm}{6cm}{\includegraphics{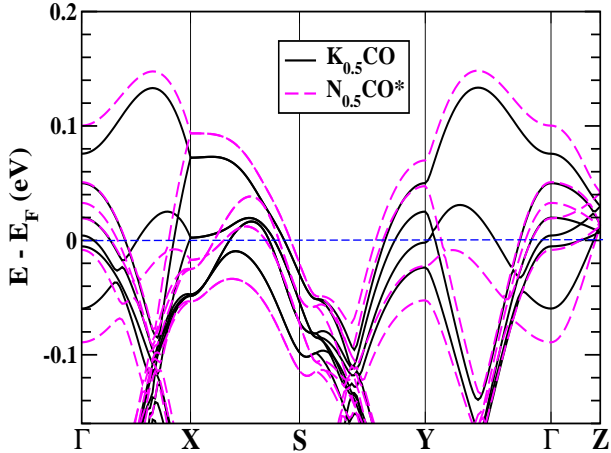}}
\caption{(Color online) Comparison of band structure
 between nonmagnetic K$_{0.5}$CoO$_2$ and Na$_{0.5}$CoO$_2^*$ near $E_F$.
 Na$_{0.5}$CoO$_2^*$ is assumed to have
 the same crystal structure as K$_{0.5}$CoO$_2$, in order to investigate
 pure effects of K substitution.
 }
\label{band3}
\end{figure}

\subsection{Identifying differences}
These differences between two cobaltates can be clarified in two ways.
First, we can determine the effects purely due to chemical
difference (K vs. Na) as opposed to the size difference leading to 
structural differences.  For this, N$_{0.5}$CO is assigned the same structure 
as in K$_{0.5}$CO and denoted N$_{0.5}$CO$^*$.
The resulting band structure enlarged near $E_F$ is compared with
that of K$_{0.5}$CO in Fig. \ref{band3}.
Even in the identical structure,  substantial differences on an important
energy scale are evident.  The top valence band in N$_{0.5}$CO$^*$ is
20 meV higher in energy at the
$\Gamma$ point, although the $t_{2g}$ bandwidth is about 5\% smaller
(not shown).  Another difference is that the projected K and Na DOS
is almost identical (and small, of course) through most of the $t_{2g}$
bands, except in a $\sim 0.15$ meV region at and below the Fermi level,
where the Na projected DOS (PDOS) is 20-35\% larger 
(more than 50\% larger at E$_F$).   
These distinctions indicate that the differences in electronic structure
are mainly due to K substitution itself rather than indirectly
through the change in structure.

Second, using the Mullikan charge decomposition, we obtained 
atom-decomposed charges, 
which are displayed in Table \ref{table2}.
The K ion is very noticeably more ionic than the Na ion, consistent
with the PDOS difference mentioned just above.  The compensating charge
is spread over the oxygen ions; the Co charges are essentially the same
for K$_{0.5}$CO and N$_{0.5}$CO$^*$.
This higher ionicity of K seems to be the most discernible difference between
these cobaltates.

\subsection{Comments on hydration}
It is still unclear what water does in the system.
The only unambiguously aspect is that hydration
dramatically increases the $c$ lattice constant, 
resulting in more two-dimensionality of the electronic system.\cite{water}
However, although the isostructural system Na$_{1/3}$TaS$_2$$\cdot y$H$_2$O 
shows very similar change in the $c$ lattice constant 
when hydrated,\cite{prb04}
$T_c \approx 4$ K in this system is independent of $y$.
This difference in behavior established that 
water has effects in the cobaltates
that are not present in the transition metal disulfides and diselenides.
In this respect it is interesting that
(Na$_{0.27}$K$_{0.12}$)CoO$_2$$\cdot$0.87H$_2$O shows
superconductivity with $T_c \approx 3$ K and 
about 7~\AA~increment in $c$ lattice constant from 
K$_{0.55}$CO, which is similar in amount to that of hydrated sodium
cobaltate.\cite{sasaki}

\section{Summary}
Using a crystal structure recently reported, we have investigated at
the LDA level the differences in electronic structure between
K$_{0.5}$CoO$_2$ and N$_{0.5}$CoO$_2$.
Comparison shows a few substantial differences
near $E_F$; smaller $t_{2g}$ bandwidth by 60 meV in K$_{0.5}$CoO$_2$,
and additional
Fermi surfaces along the $X-S$ line which are almost flat
near the $X$ point.
These differences are due more to chemical differences (higher ionic 
character of K)
rather than to structural difference between the systems.

An angle-resolved photoemission comparison of the three systems
A$_x$CoO$_2$, A = Na, K, and Rb, has appeared,\cite{arakane}
with the differences at equal doping levels being small almost too
small to quantify.
Unfortunately, samples at precisely $x$=0.5 were not the focus of
that study.   Since the superstructure we have studied is confined
to $x$=0.5, our results cannot be compared with this data.  However,
the structural disorder of the alkali at $x\neq$0.5, which extends
to the CoO$_2$ substructure, broadens the bands and hides small
distinctions.\cite{deepa}  This observation suggests that
carrying out spectroscopic studies of both systems in the insulating
phase at $x$=0.5 should be an excellent way to identify and 
characterize more precisely the effects of the different alkali cations.

\section{Acknowledgments}
We acknowledge M. D. Johannes and D. J. Singh for illuminating 
conversations, and D. Qian for clarifying the ARPES data.  
This work was supported by DOE grant DE-FG03-01ER45876 and DOE's
Computational Materials Science Network.  W.E.P. acknowledges the
stimulating influence of DOE's Stockpile Stewardship Academic Alliance Program.

\end{document}